# Graphene nanoribbons from unzipped carbon nanotubes: atomic structures, Raman spectroscopy and electrical properties

Liming Xie,[1,†] Hailiang Wang,[1,†] Chuanhong Jin,[2] Xinran Wang,[1] Liying Jiao,[1] Kazu Suenaga,[2] Hongjie Dai[1,*]

[1]Department of Chemistry, Stanford University, California 94305, USA

[2]Nanotube Research Center, National Institute of Advanced Industrial Science and Technology (AIST), Tsukuba 305-8565, Japan

[†]These authors contributed equally

Corresponding Author: hdai1@stanford.edu

**ABSTRACT:** We investigated the atomic structures, Raman spectroscopic and electrical transport properties of individual graphene nanoribbons (GNRs, widths ~10-30 nm) derived from sonochemical unzipping of multi-walled carbon nanotubes (MWNTs). Aberration-corrected transmission electron microscopy (TEM) revealed a high percentage of two-layer (2L) GNRs and some single layer ribbons. The layer-layer stacking angles ranged from $0^o$ to $30^o$ including average chiral angles near $30^o$ (armchair orientation) or $0^o$ (zigzag orientation). A large fraction of GNRs with bent and smooth edges was observed, while the rest showing flat and less smooth edges (roughness ≤1 nm). Polarized Raman spectroscopy probed individual GNRs to reveal D/G ratios and ratios of D band intensities at parallel and perpendicular laser excitation polarization ($D_{//}/D\perp$). The observed spectroscopic trends were used to infer the average chiral angles and edge smoothness of GNRs. Electrical transport and Raman measurements were carried out for individual ribbons to correlate spectroscopic and electrical properties of GNRs.

Graphene nanoribbons (GNRs) have been under intense investigation recently[1-10], with various interesting chirality, width and edge-dependent electronic properties predicted[11-13]. Theoretically, tight-binding calculations have shown that two thirds of armchair-edge GNRs are semiconducting and the other third and all zigzag-edge GNRs are metallic[11]. *Ab-initio* calculations have shown that all GNRs exhibit band gaps depending on chirality and ribbon width[12]. An interesting prediction has been reported that zigzag and chiral GNRs exhibit magnetic edge states[11, 14] and half metallicity[5], with potential applications in spintronics.

To probe interesting phenomena in GNRs, it is desirable to produce high quality materials with well defined structures in terms of chirality, width and edge structures, and characterize by atomic-scale microscopy and spectroscopy to glean structure-property relations in GNRs. Bottom-up chemical approach[2] has produced atomically smooth GNRs with width of ~1 nm and length of ~30 nm. Lithography has fabricated GNRs with disordered edges[7]. And unzipping of carbon nanotubes (CNTs) have produced GNRs with various widths and lengths up to several microns[3, 4, 6, 15]. In particular, GNRs derived in our lab by sonochemical unzipping of carbon nanotubes with minimum chemical oxidation have shown promising characteristics including smooth edges by scanning tunneling microscopy (STM)[14], the recently observed magnetic edge states in chiral GNRs[14], and low resistivity[6] compared to ribbons derived by other methods.

Here, we present the first aberration-corrected transmission electron microscopy (TEM) investigation of GNRs (widths ~10-30 nm) derived from unzipping of multi-walled carbon nanotubes (MWNTs) to probe the atomic structures of GNRs including chiral angles, edge structures and smoothness. The results were correlated with polarized Raman spectroscopy measurements to understand the relations between the chirality and edge smoothness of GNRs and their polarized Raman signatures. Further, we have performed electrical transport and polarized Raman measurements of the same individual GNRs for understanding the electrical transport and Raman spectroscopic properties of GNRs.

GNRs, produced by sonochemical unzipping of MWNTs grown by arc-discharge in an organic polymer solution[6], were deposited onto porous silicon membrane window grids for aberration-corrected TEM at an acceleration voltage of 80 or 60 kV. A large fraction of the GNRs showed Moiré patterns (Figs. 1e, 1j and Figs. 2b, 2g; coatings on the GNRs were polymer residues), indicating few-layer GNRs with non-AA/AB stacking inherited from the random stacking of concentric shells in the parent nanotubes[16, 17]. The layer number of GNRs was estimated by the number of sets of hexagonal spots in the fast Fourier transform (FFT) of the TEM images (Figs. 1f and 1k, Figs. 2c, 2h and 2l). We found that two-layer (2L) GNRs was dominant (~70%, Fig. 2m) in our sample with ~6% of monolayer (1L) GNRs (also observed by STM[14]) exhibiting a single set of hexagonal FFT spots (Figs. 2k and 2l). The layer number was confirmed for several 2L and 1L GNRs by using a focused electron-beam to ablate and remove carbon atoms[18] from GNRs layer by layer (Fig. S3 in Supporting Information [SI]). Also, due to the coexistence of nanotubes in the sample, we confirmed the flat nature (as opposed to cylindrical shape) of several ribbons by tilting the sample stage relative to the electron beam and observing reduced, projected widths of the ribbons (Fig. S4 in SI).

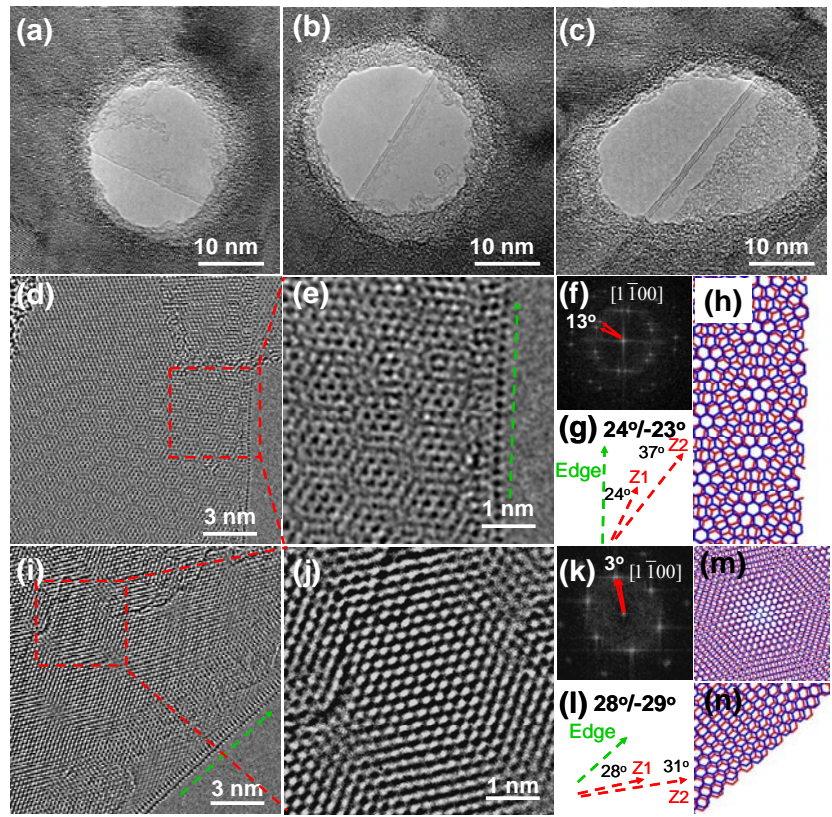

**Figure 1.** TEM images of GNRs with straight edge lines. (a), (b), (c) Low-magnification TEM images of GNRs with one, two and three straight edge lines, respectively. Polymer residues are visible on the ribbons. (d) Large-scale TEM image, (e) zoomed-in TEM image, (f) FFT image, and (g) chiral angle analysis of a GNR with one straight edge line. (h) An atomic model of the GNR in panel (d) and (e). (i) Large-scale TEM image, (j) zoomed-in TEM image, (k) FFT image, and (l) chiral angle analysis of a GNR with two straight edge lines. (m), (n) Atomic models of the GNR inside the GNR and near the edge respectively. The green dashed lines indicate GNR axial direction. The zigzag directions for each layer (Z1 and Z2, indicated by red dashed lines) for the different layers in panels (g) and (l) are derived from the $[1\bar{1}00]$ directions (indicated by red arrows) in the corresponding FFT images.

The chiral angle of each layer in a GNR was determined by measuring the angle between the GNR axial direction and the zigzag direction (i.e., the $[1\bar{1}00]$ direction) of each layer from the FFT. We defined the zigzag direction to be at $0^o$ chiral angle. Counter-clockwise direction from the zigzag line was defined to have positive chiral angles. While most of the GNRs showed random chiral angles for each layer (Fig.1d-1h for a 2L GNR with $24^o/-23^o$ chiral angles, and chiral angle distribution in Fig. S5a in SI), we did observe a small fraction of 2L GNRs with both layers oriented close to the armchair directions (Fig. 1i-1n for a GNR with $28^o/-29^o$ chiral angles) or close to the zigzag directions (Fig. S6 in SI for a GNR with $0^o/8^o$ chiral angles). The layer-layer stacking for 2L GNRs, measured from the rotation angle between the hexagonal spot sets in the FFT images, also ranged randomly from $0^o$ to $30^o$ (Fig. S5b in SI).

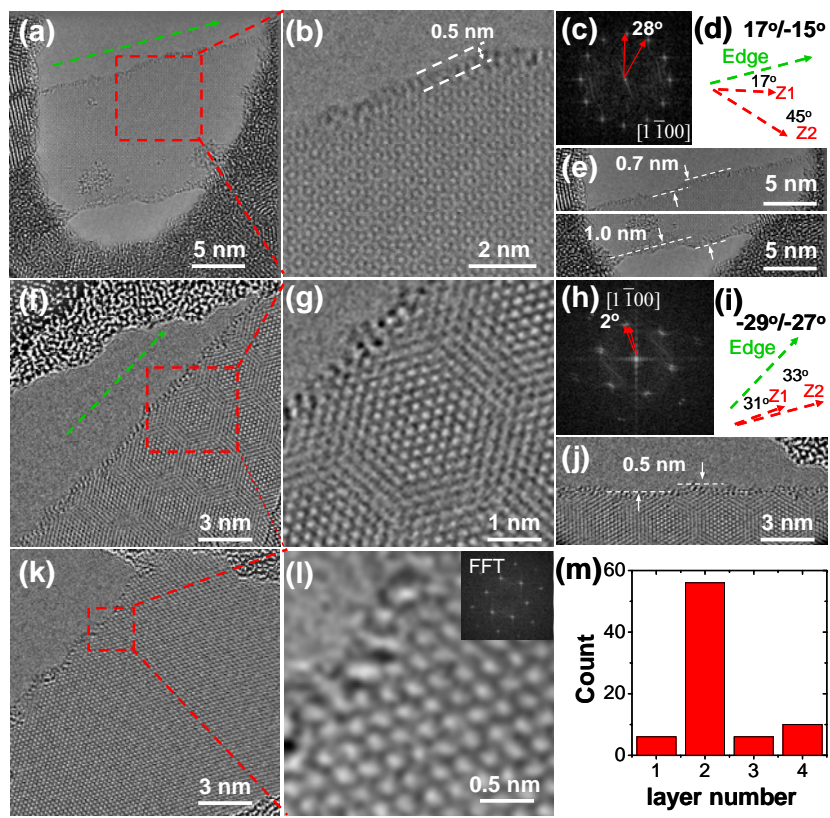

**Figure 2.** TEM images of GNRs with flat edges and distribution of GNR layer numbers. (a), (f) Large-scale TEM images, (b), (g) zoom-in TEM images, (c), (h) FFT images, (d), (i) chiral angle analysis, (e), (j) edge smoothness analysis of 2L GNRs with flat edges. The green dashed lines indicate GNR axial directions. The zigzag directions (Z1 and Z2, indicated by red dash arrows) for the different layers in panel (d) and (i) are derived from the $[\bar{1}100]$ directions (indicated by red arrows) in the corresponding FFT images. (k) Large-scale TEM image, (l) zoomed-in TEM image of a 1L GNR. (m) Distribution of layer numbers for GNRs in the sample.

TEM imaging revealed that many GNRs (66 out of 85) showed straight, parallel edge lines (Figs. 1a-1c). The dark, straight edge lines of the GNRs suggested likely bending at the edges, similar to bent edges observed in few-layer graphene sheets[19]. The edges of such ribbons appeared very smooth over relatively long ribbon lengths (straight edge lines in Figs. 1a-1c, 1d and 1i), although the edge bending made it difficult to discern possible roughness out of the ribbon plane. We also observed GNRs with flat edges (19 out of 85) but without the dark parallel edge lines (Figs. 2a, 2f and 2k). The edges of these ribbons tended to be less smooth with an edge roughness on the order of ~1 nm (see Fig. 2a, 2b and 2e for a GNR with edge roughness ~1 nm over ~20 nm length). A flat-edge GNR with relatively smoother edges is shown in Fig. 2f, 2g and 2j (edge roughness <0.5 nm). For all of the GNRs imaged, we observed few obvious defects or disorders inside the GNR plane (Figs. 1d and 1i, Figs. 2a, 2f and 2k), indicating high quality of the GNRs.

In general, TEM revealed that the GNRs exhibited few defects in the plane, smooth edges (edge roughness ≤1nm), random stacking between layers and various chiral angles including GNRs with average layer orientations near armchair or zigzag directions. Next, we used polarized micro-Raman spectroscopy to characterize individual GNRs on $SiO_2$/Si substrates in the 'VV' configuration (i.e., with the excitation laser polarization parallel to the polarization of the detected Raman signal; see Fig. S2 in SI). For polarized Raman measurements, we used AFM imaging to select a relatively small percentage of GNRs with apparent topographic heights in the lowest range of 1.0-1.2 nm (including polymer residues on the ribbon) in the sample. These ribbons were likely 1L GNRs though the possibility of 2L could not be ruled out.

We observed four Raman bands for individual GNRs, including the disorder related D band at ~1350 cm$^{-1}$, the graphitic G band at ~1600 cm$^{-1}$, the D' band at ~1620 cm$^{-1}$ (a disorder related intravalley double-resonance Raman band) and the 2D band at ~2700 cm$^{-1}$ (or G' band, corresponding to an intervalley double-resonance Raman band) (Figs. 3a-3c). Since the layers in GNRs from unzipped MWNTs were non-AB stacked, we were unable to determine the layer number of our GNRs accurately based on the 2D profile. We found that the D and D' bands of GNRs exhibited high polarization dependence, reaching maximum (or minimum) intensities when the laser-Raman polarization was parallel (or perpendicular) to the GNR direction (Fig. 3a-3c). On the other hand, the G and 2D bands showed weak polarization dependence for the 10-30 nm wide GNRs, which differed from CNTs. The G band intensity approached zero at perpendicular laser polarization for individual SWNTs[20], few-walled CNTs (Fig. S8 in SI) and our GNR's parent MWNTs (Fig. S9 in SI) measured in our control experiments.

With few defects in the plane of GNRs (TEM data in Figs. 1 and 2), the observed D bands of the GNRs were likely due to the edges. Previous Raman spectroscopy of graphene edges showed high D/G ratios for near armchair edge orientations (chiral angles near $30^o$) due to favoring the intervalley resonance Raman processes responsible for the D band[21-24]. No D band or weak D band was expected for smooth zigzag edges due to the disfavored intervalley resonance for zigzag oriented edges. Therefore, we tentatively used the $D_{//}/G_{//}$ ratio (D band intensity over G band intensity at parallel polarization) to infer the chiral angle (or averaged chiral angles for possible 2L GNRs selected for Raman experiments) of the ribbons. For more than 20 GNRs measured, the $D_{//}/G_{//}$ ratio spanned a large range from 0.4 to 9.5 (a near 25-fold variation, Fig. 3e). GNRs with high $D_{//}/G_{//}$ ratio near 9.5 (Fig. 3a) were assigned to GNRs with average chiral angles close to 30° (near armchair orientation) and GNRs with low $D_{//}/G_{//}$ ratios around 0.4 (Fig. 3b and 3c) were assigned to GNRs with average chiral angles close to 0° (near zigzag orientation). GNRs with intermediate $D_{//}/G_{//}$ ratios (Fig. S10 in SI) were assigned to chiral angles in between 0° and 30°.

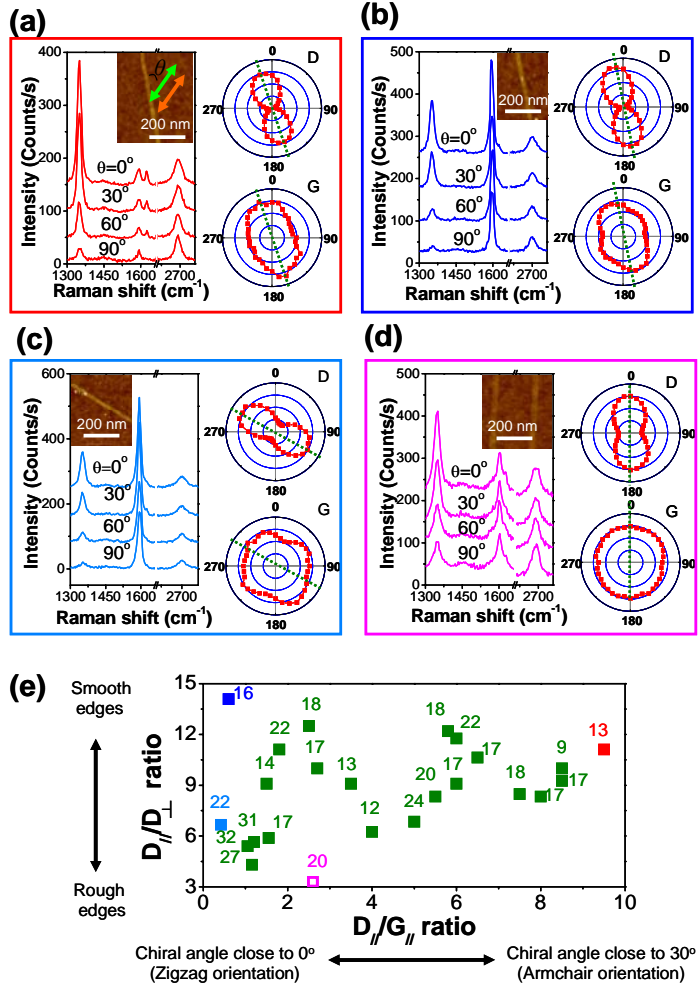

**Figure 3.** Polarized Raman spectra of GNRs (AFM height ~1.0-1.2nm). (a) An individual GNR exhibiting a high $D_{//}/G_{//}$ ratio of 9.5 and a high $D_{//}/D_{\perp}$ ratio of 11. (b) An individual GNR exhibiting a low $D_{//}/G_{//}$ ratio of 0.6 and a high $D_{//}/D_{\perp}$ ratio of 14. (c) An individual GNR exhibiting a low $D_{//}/G_{//}$ ratio of 0.4 and a low $D_{//}/D_{\perp}$ ratio of 6.7. (d) Data for lithographically patterned GNRs. The insets show AFM images for the corresponding GNRs. The intensity scales in all polar plots are linear starting from zero. The green dash lines indicate the GNR axial directions. Polarization dependence of D' and 2D band intensities is shown in Fig. S7 in SI. (e) $D_{//}/D_{\perp}$ ratio vs. $D_{//}/G_{//}$ ratio for all measured GNRs. The numbers near the data points are the corresponding GNR widths. The hollow square point represents the data for the Litho-GNRs in panel (d). The red, blue, and cyan points are the data for the GNRs in panels (a)-(c) with spectra in the corresponding colors.

The intensity ratio of the D band at parallel polarization ($D_{//}$) and perpendicular polarization ($D\perp$) measured at graphene edges have been suggested to reflect edge roughness and edge chirality[21]. A rougher edge generally exhibits lower $D_{//}/D\perp$ since the existence of disordered segments and random-orientated armchair segments can significantly increase $D\perp$ intensity and hence lower the $D_{//}/D\perp$ ratio. The $D_{//}/D\perp$ ratios of near-zigzag (or near-armchair) edges are strongly (or weakly) dependent on edge roughness[21]. In our case, for GNRs with high $D_{//}/G_{//}$ ratios (~8-9.5, tentatively assigned to average chiral angle near 30° or near armchair orientation), high $D_{//}/D\perp$ ratios (> ~8) in a narrow range was typically observed (Fig. 3e). For GNRs with low $D_{//}/G_{//}$ ratios (average chiral angle close to 0° or near zigzag orientation), the observed $D_{//}/D\perp$ ratios spanned a much wider range from ~ 14 down to ~4 (Fig. 3e). These results were supportive of our assignment of GNRs with high (or low) $D_{//}/G_{//}$ ratios to ribbons with averaged orientations of the layers near the armchair (or zigzag) direction.

For comparison, we fabricated lithographically patterned GNRs (Litho-GNRs) with width of ~20 nm[25] (Fig. 3d). The edges were known to be rough and disordered, causing transport gaps in Litho-GNRs observed experimentally[7]. Polarized Raman measurements of the Litho-GNRs found $D_{//}/D\perp$ ratios of ~3-4 (Figs. 3d and the hollow square in Fig. 3e), obviously lower than $D_{//}/D\perp$ ratios of GNRs derived from nanotube unzipping. This spectroscopically confirmed that GNRs from unzipped MWNTs were of higher edge quality than Litho-GNRs.

We found that both $D_{//}/G_{//}$ and $D_{//}/D\perp$ ratios showed discernable increasing trends as the GNR width decreased in the 30 to 10 nm range (Fig. S11 in SI), which is consistent with the measured electron coherence length of ~3 nm[26] near graphene edges and also the measured D/G ratio for Litho-GNRs by another group[27]. Different from few-layer graphene with non-AA/AB stacking[28], the 2D/G intensity ratio can not be used to indicate layer number for GNRs because $2D_{//}/G_{//}$ ratio, as well as 2D width, was dependent on $D_{//}/G_{//}$ ratio (Fig. S12 in SI). Theoretically, it was suggested that, for GNRs[29, 30] and graphene sheet edges, the polarization dependence of G band intensity was different for zigzag and armchair edges[31]. However, the measured polarization dependence of the G band of our GNRs (Figs. 3a-3d) did not match with theoretical calculations[29, 30], which could be due to much wider ribbons measured here than in theoretical calculations. Further investigations are required to understand the differences.

We made electrical contacts to some of the same individual GNRs with different $D_{//}/G_{//}$ ratios and $D_{//}/D\perp$ ratios as measured by micro-Raman. Electrical transport measurements found no obvious dependence of GNR resistivity (defined as resistance at the Dirac point × GNR width/length) on $D_{//}/G_{//}$ ratio (Fig. 4e). Two GNRs with very different $D_{//}/G_{//}$ ratio of 6.5 and 0.6 (Fig. 4a and 4b with similar $D_{//}/D\perp$) exhibited similar resistivity (see red and blue squares in Fig. 4e). This was consistent with that the $D_{//}/G_{//}$ ratio reflected average lattice orientation rather than defects in the GNRs. On the other hand, for two GNRs (Fig. 4b and 4d) with similar average chiral angles near 0° ($D_{//}/G_{//}$ ratio ~ 0.6 and 0.4 respectively), the GNR with a higher $D_{//}/D\perp$ ratio of 14 (Fig. 4b and the blue square in Fig. 4e) exhibited lower resistivity than the GNR with a lower $D_{//}/D\perp$ ratio of 6.7 (Fig. 4d and the cyan square in Fig. 4e). Similarly, for two GNRs with chiral angles closer to 30° (Figs. 4a and 4c, $D_{//}/G_{//}$ ratios of 6.5 and 7.5, respectively), the GNR with a higher $D_{//}/D_\perp$ ratio also exhibited a lower resistivity. For various GNRs measured, a discernable trend of lower resistivity for GNRs with higher $D_{//}/D_\perp$ ratios existed (Fig. 4f), consistent with reduced edge scattering in GNRs with smoother edges (higher $D_{//}/D_\perp$).

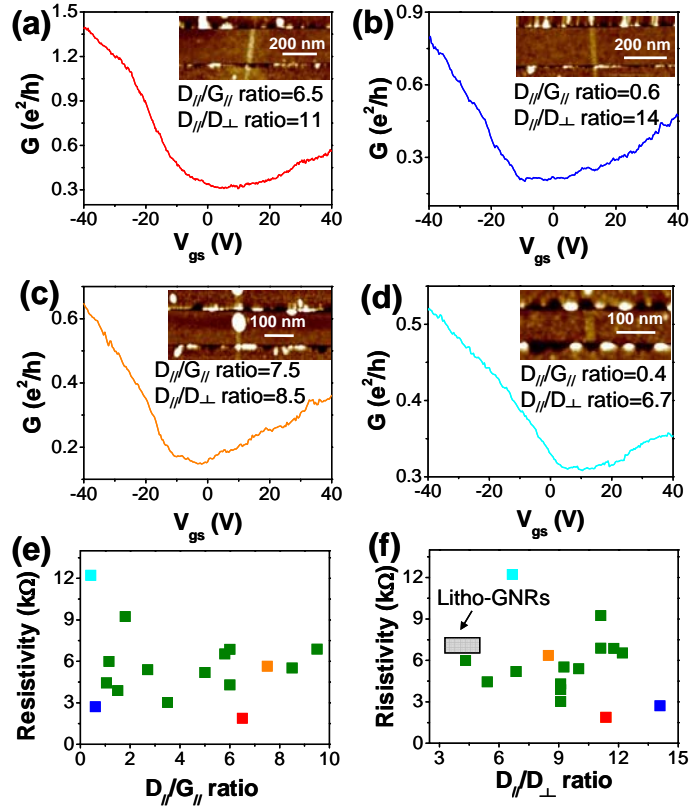

**Figure 4.** Room-temperature electrical transport measurements of GNRs used in polarized Raman measurements. (a)-(d) Conductance vs. gate voltage ($V_{gs}$) for four GNRs with $D_{//}/G_{//}$ and $D_{//}/D_\perp$ ratios indicated. The source-drain bias voltage ($V_{ds}$) was 10 mV. The insets are the AFM images of the corresponding GNR devices. (e) Plot of GNR resistivity vs. Raman $D_{//}/G_{//}$ ratio for various GNRs measured. (f) Plot of GNR resistivity vs. $D_{//}/D_\perp$ ratio for various GNRs. The red, blue, orange and cyan points in panels (e) and (f) are data points for the GNRs in panels (a)-(d) with curves in the corresponding colors. Note that we used AFM imaging to select a relatively small percentage of GNRs with apparent topographic heights in the lowest range of 1.0-1.2 nm (including polymer residues on the ribbon) for Raman and transport measurements. These ribbons were likely 1L GNRs with an average resistivity higher than that of 1-3 nm tall GNRs (a high percentage of two-layer ribbons) measured previously (ref. 6). Raman mapping for locating GNRs before polarized Raman measurements might also decrease the conductance of the GNRs measured in this paper.

In conclusion, atomic-scale TEM imaging was done for high quality GNRs from unzipped MWNTs to reveal layer numbers, layer stacking, average chiral angles and edge smoothness. The results were combined with polarized Raman to suggest that GNRs with $D_{//}/G_{//}$ ratio in the range of 0.4 to 9.5 corresponded to average chiral angles from 0° (zigzag oriented) to 30° (armchair oriented). GNRs with decreasing $D_{//}/D_\perp$ ratio (in the range of 4-14) corresponded to lower degree of edge smoothness (with lithography derived GNRs exhibiting $D_{//}/D_\perp$ ratio below 4), and the inferred edge roughness was consistent with electrical transport measurements of GNRs.

**SUPPORTING INFORMATION**

Materials and methods, including sample preparation, TEM characterizations, AFM characterizations, polarized Raman measurements, electrical devices and measurements; electron-beam (e-beam) layer-by-layer evaporation of GNRs; TEM images of a GNR with straight edge lines at different tilting angles; layer-layer stacking in 2L GNRs and chiral angles for all GNRs; TEM images of a 2L GNR with chiral angles of $0°/8°$; AFM images and polarization dependence of D' and 2D band intensities for the GNRs in Figs. 3a-3d; polarized Raman measurements on individual few-walled CNTs; polarized Raman measurements on individual MWNTs (the parent material used in unzipping experiments for obtaining GNRs used in this work); polarized Raman spectra of a GNR with an intermediate $D_{//}/G_{//}$ ratio of 2.7; dependence of $D_{//}/G_{//}$ ratio and $D_{//}/D_{\perp}$ ratio on GNR width; $2D_{//}/G_{//}$ ratio and 2D width vs. $D_{//}/G_{//}$ ratio for GNRs from unzipped MWNTs.

**ACKNOWLEDGMENT**

This work was support by ONR, Graphene-MURI, MARCO-MSD, Intel and the NCEM at Lawrence Berkeley Lab, which was supported by the US Department of Energy under Contract # DE-AC02-05CH11231. H.W. acknowledges Bin Jiang from FEI, Chengyu Song and Peter Ercius from NCEM for training and supervision on TEAM0.5 imaging. C.J. and K.S. acknowledge the support of JST-CREST.

# Supporting information

# Graphene nanoribbons from unzipped carbon nanotubes: atomic structures, Raman spectroscopy and electrical properties

Liming Xie,[1,†] Hailiang Wang,[1,†] Chuanhong Jin,[2] Xinran Wang,[1] Liying Jiao,[1] Kazu Suenaga,[2] Hongjie Dai[1,*]

[1]*Department of Chemistry, Stanford University, California 94305, USA*

[2]*Nanotube Research Center, National Institute of Advanced Industrial Science and Technology (AIST), Tsukuba 305-8565, Japan*

[†]These authors contributed equally to this work.

[*]To whom correspondence should be addressed: hdai1@stanford.edu

## Part 1. Materials and Methods

### Sample Preparation

GNRs were produced by unzipping arc-discharge grown multi-walled carbon nanotubes (MWNTs)[1]. MWNTs (30 mg; Aldrich, 406074-500MG) were calcined at 500 °C in a furnace for 1 h. The calcined nanotubes (15 mg) and 7.5 mg PmPV (Aldrich, 555169-1G) were then dissolved in 10 ml 1,2-dichloroethane and sonicated (Cole Parmer sonicator, model 08849-00) for 1 h. The solution was ultracentrifuged (Beckman L8-60M ultracentrifuge) at 40,000 rpm for 2 h. For TEM sample preparation, silicon membrane window grids (SPI Supplies, US200-P15Q UltraSM 15 nm Porous TEM Windows) were soaked in the GNR solution overnight. The grid with GNRs deposited was calcined in air at 350 °C for 30 min to remove polymer coating on the GNRs to a small extent.

GNRs on $SiO_2$/Si substrates ($SiO_2$ thickness of 300 nm) were prepared by spin-coating a GNR suspension on $SiO_2$/Si substrates followed by calcination in air at 350 °C for cleaning. The samples were used for AFM, micro-Raman measurements and/or electrical device fabrication. For GNRs characterized by Raman spectroscopy and followed by device fabrication, relatively long GNRs (~2 $\mu$m) with the lowest topographic heights (in the range of 1.0-1.2 nm) were selected. Raman mapping was done near one end of the ribbon and devices were fabricated on the other end of the ribbon without laser irradiation.

**TEM Characterizations**

Low-resolution TEM was done on an FEI Tecnai G2 F20 X-TWIN TEM instrument at an operation voltage of 200 kV at Stanford University. High-resolution, aberration-corrected TEM was done at an operation voltage of 80 kV on TEAM 0.5 [2] at the Lawrence Berkeley National Laboratory and at an operation voltage of 60 kV on a JEOL 2100F TEM with the DELTA correctors[3] (including an imaging aberration corrector) at the Nanotube Research Center, National Institute of Advanced Industrial Science and Technology (AIST) in Japan.

**AFM Characterizations**

AFM was done on a Vecco IIIa nanoscope using the tapping mode. Tip convolution was calibrated in the width measurements of GNRs using the same method in the reference[4]. AFM was used to locate GNRs relative to pre-fabricated alignment markers and the registered ribbons were used for subsequent micro-Raman and device fabrication experiments. In detail, large scale AFM image, such as 12 $\mu$m by 12 $\mu$m, was taken with maker and GNRs included in the image (Figure S1a). The position of the individual GNRs were measured relatively to the maker. After that, the sample was put under Raman microscope, the marker was directly seen under optical microscope and then the locations of individual GNRs were derived. Raman mapping was used to further confirm the GNRs and also to image the GNR axis directions (Figure S1c).

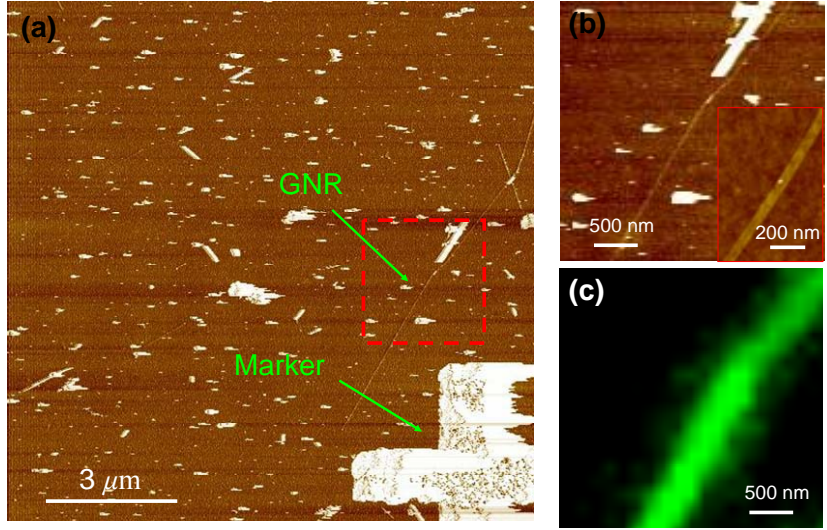

**Figure S1** (a) Large scale AFM image (phase image) of GNRs on SiO$_2$/Si substrate. Relative position of the GNRs to the maker can be determined from the AFM image. (b) Zoomed-in AFM images and (c) Raman mapping (D band intensity) of the area indicated by the red dash box in panel a. The inset in (b) shows a high resolution AFM image of the GNR.

**Polarized Raman Measurements**

Polarized Raman measurements were done on a Horiba HR800 Raman system with 532 nm excitation. The laser power was kept at ~1 mW/$\mu m^2$ during Raman measurements. A 100x objective and a 300 lines/mm grating was used, corresponding to a spatial resolution of ~0.5 $\mu$m and a spectrum resolution of ~2 cm$^{-1}$. A half-wave plate was put in the laser path to rotate the polarization of the laser (Fig. S2a). An analyzer was put in the signal path to select the polarization of the Raman signal (Fig. S2a). The GNR direction (with an accuracy of +/- 5$^o$) was measured by both AFM imaging and Raman mapping.

Polarized Raman measurements were done in the *VV* configuration (Fig. S2b), in which the polarization directions of the laser and the Raman signal were kept parallel and the Raman spectra were collected at different angles between the laser-Raman polarization direction and the GNR direction. Typical Raman accumulation time was 14 s. Polarization dependent response of the Raman system was calibrated by using Raman bands of graphite. In detail, an analyzer was put in the signal path to yield polarized Raman signal and a half-wave plate was put after the analyzer. Polarization dependent response of the Raman system was recorded while rotating the half-wave plate. The intensities of D band, G band, D' band and 2D (or G') band were extracted through Lorentzian peak fitting of the Raman spectra.

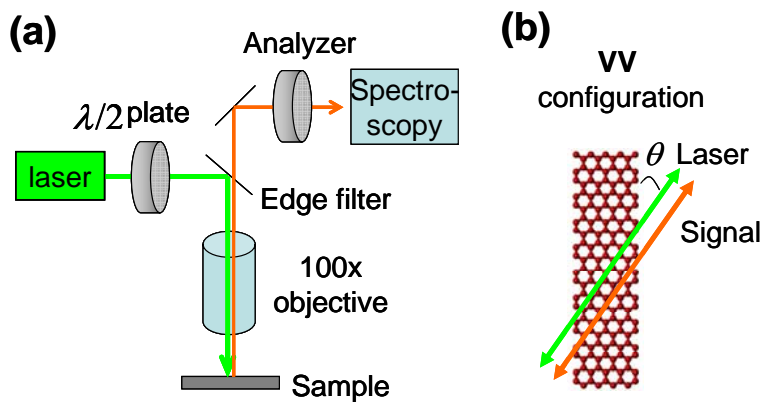

**Figure S2** (a) Scheme of polarized Raman measurements. (b) Illustration of *VV* configuration. The green and orange arrows show the polarization directions of the laser and the analyzer, respectively. θ is the angle between the laser-Raman polarization direction and the GNR axis direction.

**Electrical Devices and Measurements**

Electrical devices were made for the same GNRs (at locations determined by AFM imaging) used in micro-Raman mapping experiments by electron-beam lithography followed by electron-beam evaporation of Pd (30 nm). The devices were annealed in Ar at 220 °C for 15 min to improve the contacts. Electrical measurements were done at room temperature in a home-built vacuum probe station and after electrical annealing[5] to clean the ribbons and observe the Dirac point of the GNR devices. The heavily doped Si substrate was used as a gate. The resistivity of an individual GNR, $R$, was calculated by $R = \left(\frac{V_{ds}}{I_{ds}}\right)_{V_{gs}, Dirac\ point} \frac{W}{L}$, where $V_{ds}$ is the drain-source voltage, $I_{ds}$ is the drain-source current, $V_{gs,\ Dirac\ point}$ is $V_{gs}$ at the minimum conductance point, $W$ is the width of the GNR in nm and $L$ is the channel length of the GNR device in nm.

**Part 2. Electron-beam (e-beam) layer-by-layer evaporation of GNRs.**

The high energy electrons used in the TEM imaging can evaporate carbon atoms layer by layer[6], which can be used to count the layer numbers of GNRs.

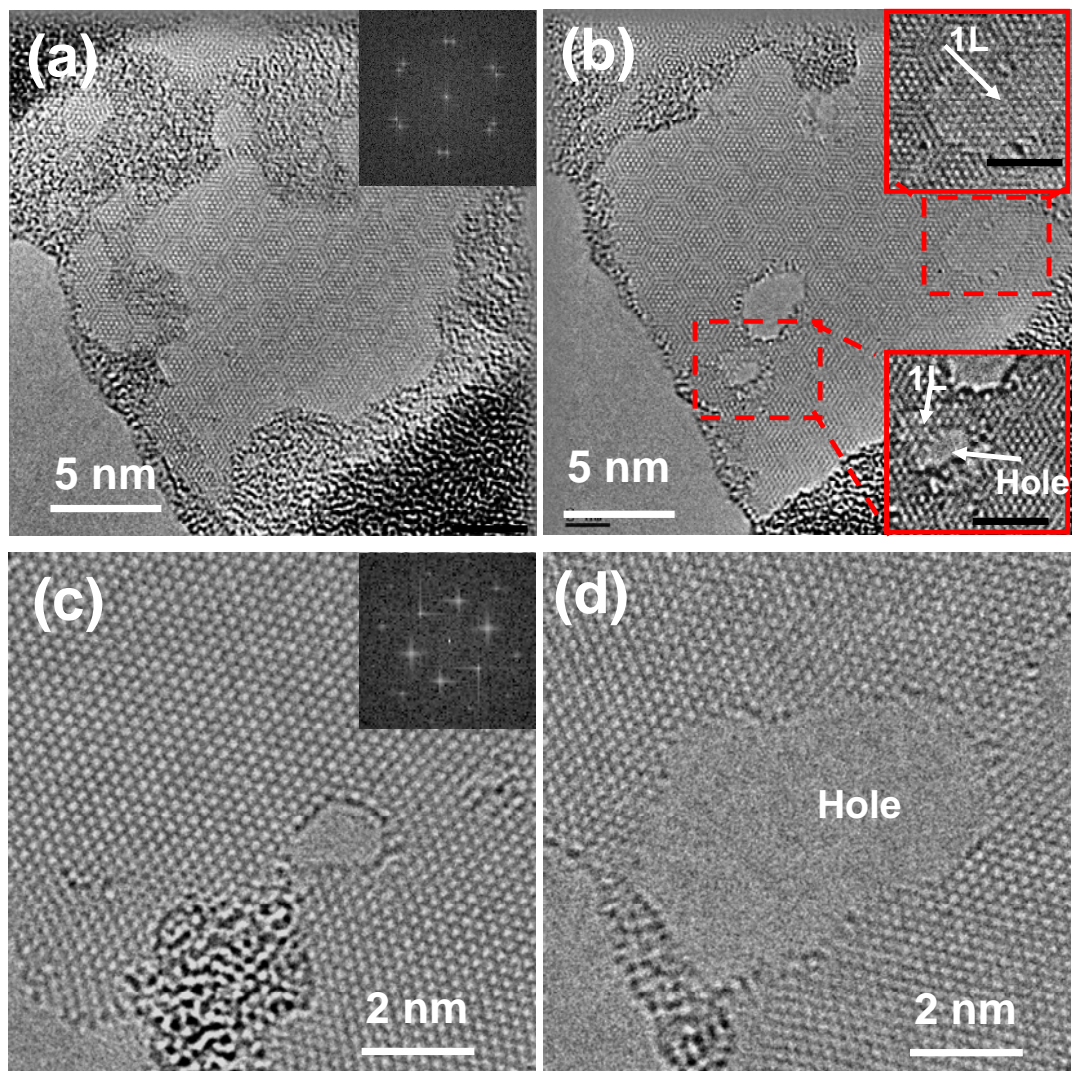

**Figure S3** A GNR with two sets of hexagonal spots in the FFT image (a) before and (b) after a few minutes of e-beam irradiation. Coatings on the ribbons are polymer residues from GNR making process. The scale bars in panel (b) insets are 2 nm. A GNR with one set of hexagonal spots in the FFT image (c) before and (d) after e-beam irradiation. The insets in panels (a) and (c) are FFT images of the corresponding GNR regions in panels (a) and (c). Note: the simple FFT analysis is not always reliable to determine the layer number, because one could underestimate it if there is any well-ordered AA or AB stacking involved. Therefore it is essential to confirm the layer number by layer-by-layer ablation using a focused e-beam.

**Part 3. TEM images of a GNR with straight edge lines at different tilting angles.**

Figure S4 shows a GNR with straight edge lines at tilting angles of $0°$ and $23°$ with widths of 20.7 and 19.3 nm, respectively. This result is consistent with a model of tilting a flat ribbon. In detail, for a flat ribbon with a width of 20.7 nm and an angle of $14°$ from the tilting angle, the projected width at a tilting angle of $23°$ should be

$$\sqrt{[20.7 \bullet \sin(14^o)]^2 + [20.7 \bullet \cos(14^o) \bullet \cos(23^o)]^2}\ nm\ =\ 19.2\ nm$$

The observed width (19.3 nm) is 0.1 nm from the calculation, which is within the imaging resolution (~0.1 nm).

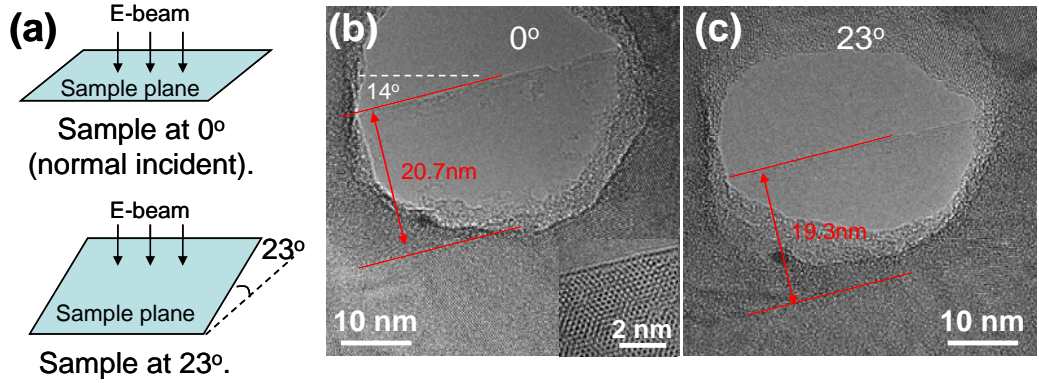

**Figure S4** (a) Schematic illustration of imaging GNR sample at different titling angles. TEM of a GNR with straight edge lines at tilting angles of (b) $0^o$ and (c) $23^o$. The GNR is $14^o$ from the tilting axis [the white dash line in panel (b)]. Inset in (b) shows a high resolution image of this GNR at tilting angle of $0^o$.

**Part 4. Layer-layer stacking in two-layer GNRs and chiral angles for all GNRs.**

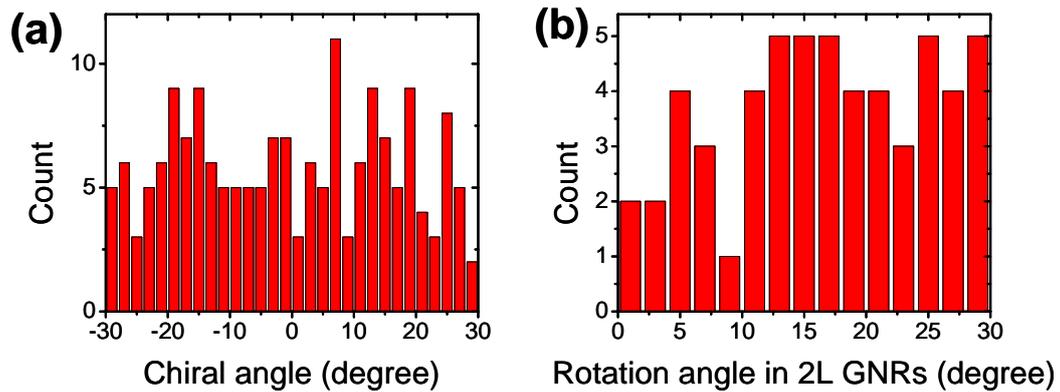

**Figure S5** (a) Distribution of chiral angles for all imaged GNRs. (b) Distribution of layer-layer stacking in 2L GNRs.

**Part 5. TEM images of a two-layer GNR with chiral angles of $0^o/8^o$.**

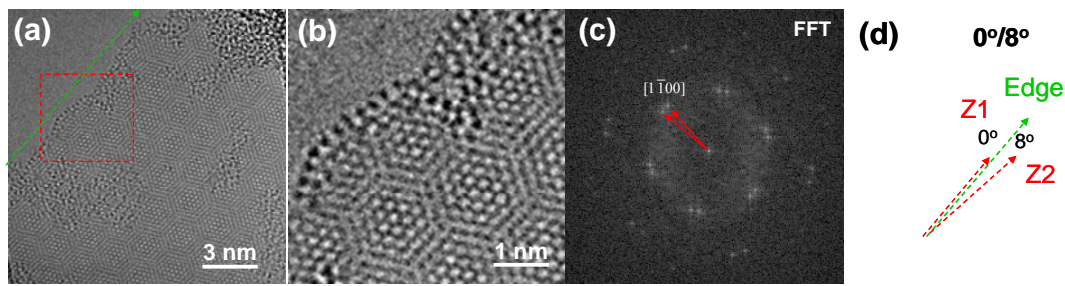

**Figure S6** (a) Large-scale and (b) zoom-in TEM images of a 2L GNR with chiral angles of 0°/8°. (c) FFT image of this two-layer GNR. The green dash arrows indicate the edge direction. The red arrows indicate the $[1\bar{1}00]$ directions for the two layers. (d) Chiral angle analysis. The zigzag directions (Z1 and Z2, indicated by red dash arrows) for the two layers are derived from the $[1\bar{1}00]$ directions in panel (c).

**Part 6. AFM images and polarization dependence of D' and 2D band intensities for the GNRs in Figs. 3a-3d.**

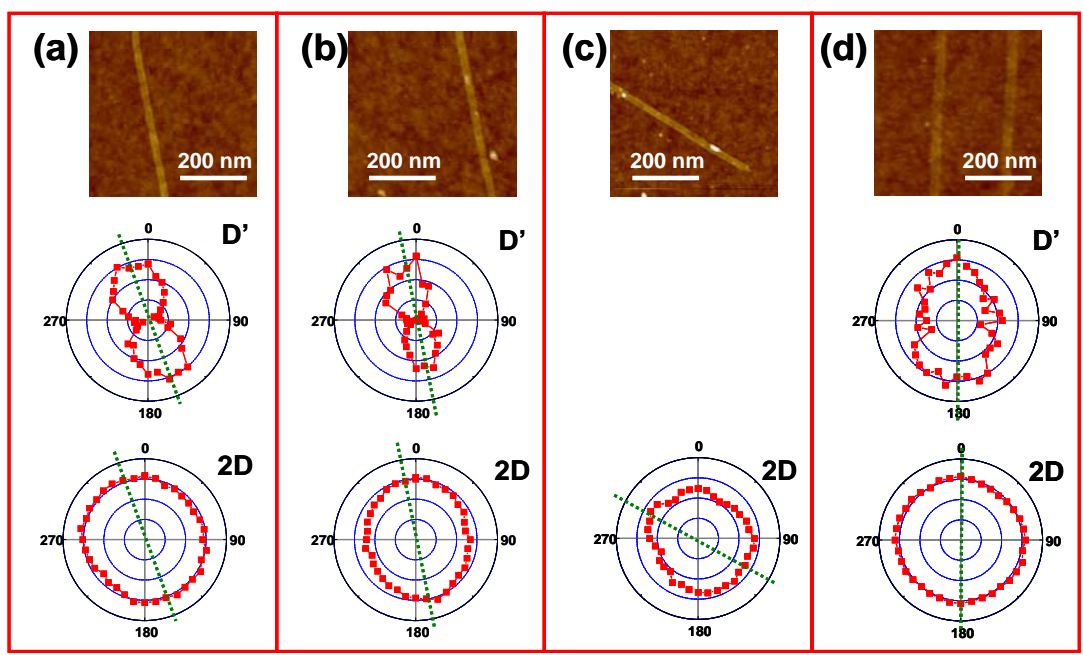

**Figure S7** AFM images and polarization dependence of D' and 2D band intensities for the GNRs in (a) Fig. 3a, (b) Fig. 3b, (c) Fig. 3c and (d) Fig. 3d. The intensity scales in all polar plots are linear from zero. The green dash lines indicate the GNR directions. The D' band intensity for the GNR in panel (c) was too weak to be extracted out. The intensity scales in all polar plots are linear from zero. The green dash lines indicate the GNR directions.

**Part 7. Polarized Raman measurements on individual few-walled CNTs.**

Few-walled CNTs (mainly double-walled CNTs) were synthesized by chemical vapor deposition[7] and was a gift from Dr. Jie Liu at Duke University.

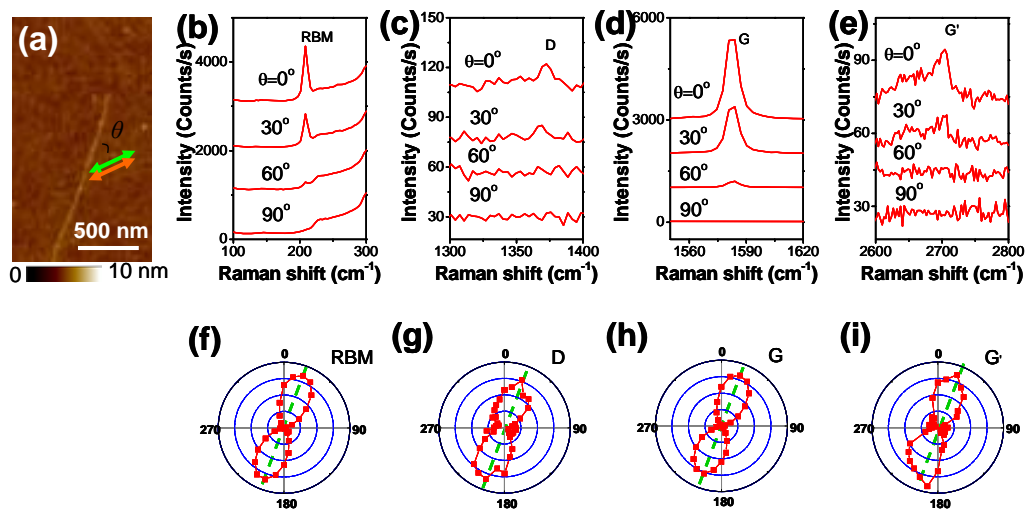

**Figure S8** (a) AFM image of a few-walled CNT. Polarized Raman spectra of this CNT at different laser-Raman polarization to nanotube axis angles: (b) RBM band, (c) D band, (d) G band, and (e) G' band. Polarization dependent Raman intensity for (f) RBM band, (g) D band, (h) G band, and (i) G' band. The intensity scales in all polar plots are linear from zero. The green dash lines indicate the GNR direction.

**Part 8. Polarized Raman measurements on individual MWNTs (the parent material used in unzipping experiments for obtaining GNRs used in this work).**

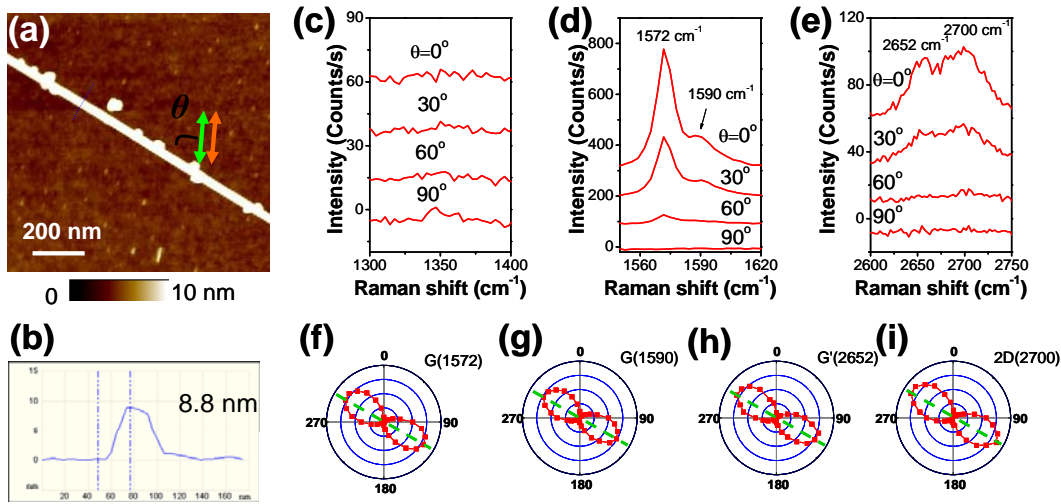

**Figure S9** (a) AFM image of a MWNT. (b) Section analysis of this MWNT along the blue dash line in panel (a), showing a height of 8.8 nm. Polarized Raman spectra of this carbon nanotube at different laser-Raman polarization to nanotube axis angles: (c) 1300-1400 cm$^{-1}$ range, (d) G band, and (e) G' band. Polarization dependent Raman intensity for (f) G band at 1572 cm$^{-1}$, (g) G band at 1590 cm$^{-1}$, (h) G' band at 2652 cm$^{-1}$, and (i) G' band at 2700 cm$^{-1}$. The intensity scales in all polar plots are linear from zero. The green dash lines indicate the GNR direction.

**Part 9. Polarized Raman spectra of a GNR with an intermediate $D_{//}/G_{//}$ ratio of 2.7.**

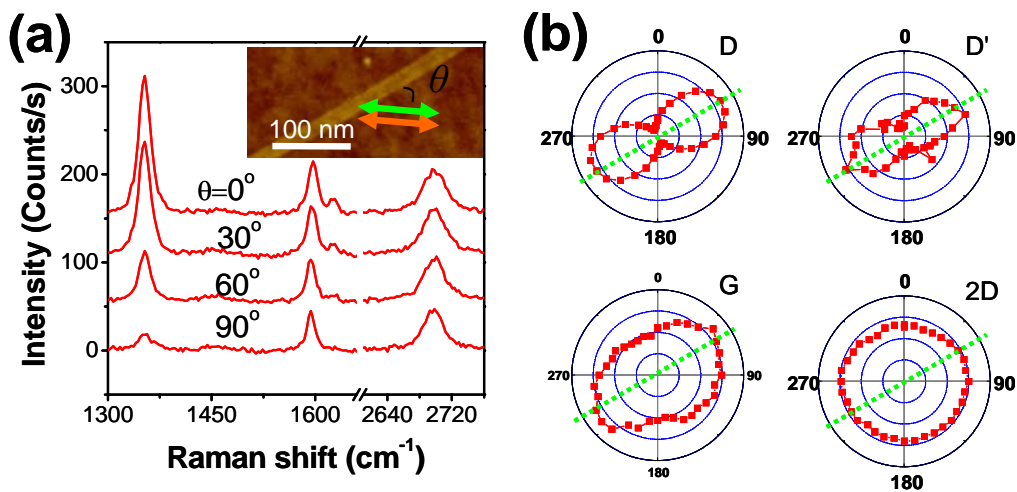

**Figure S10** (a) Polarized Raman spectra of a GNR with an intermediate $D_{//}/G_{//}$ ratio of 2.7. The inset shows an AFM image of the GNR. (b) Plots of D, D', G and 2D band intensities to the angle between the laser-Raman polarization direction and the GNR direction. The intensity scales in all polar plots are linear from zero. The green dash lines indicate the GNR direction.

## Part 10. Dependence of $D_{//}/G_{//}$ ratio and $D_{//}/D_{\perp}$ ratio on GNR width.

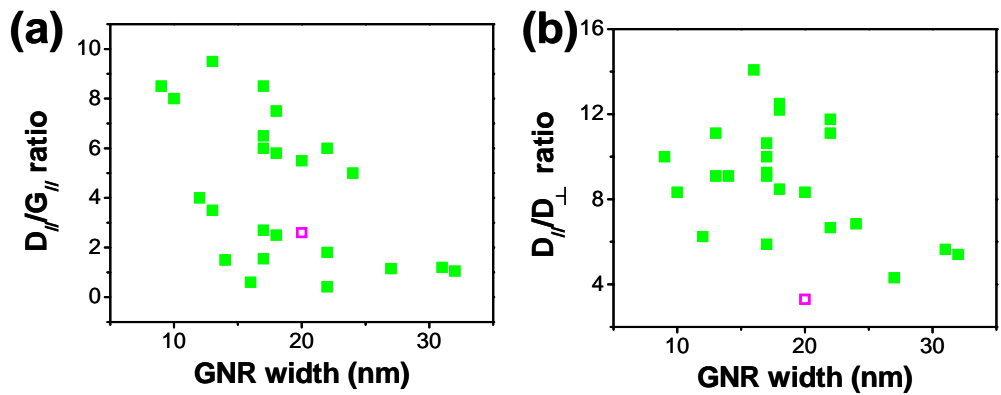

**Figure S11** Dependence of (a) $D_{//}/G_{//}$ ratio and (b) $D_{//}/D_{\perp}$ ratio on GNR width. The hollow square is the data for Litho-GNRs in Fig. 3d.

## Part 11. $2D_{//}/G_{//}$ ratio and 2D width vs. $D_{//}/G_{//}$ ratio for GNRs from unzipped MWNTs.

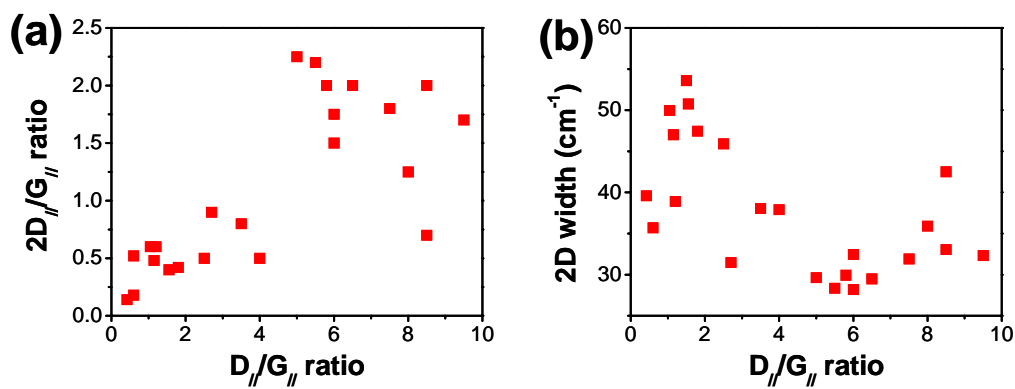

**Figure S12** (a) $2D_{//}/G_{//}$ ratio and (b) 2D width vs. $D_{//}/G_{//}$ ratio for GNRs from unzipped MWNTs.